%% file: compilethisfile.tex
\documentclass[final,5p,times,twocolumn,authoryear]{elsarticle}

\usepackage{amssymb}

\usepackage{lineno}

\usepackage{url}
\usepackage{csvsimple}
\usepackage{float}
\usepackage{colortbl}
\newcommand{\commentblock}[1]{}
\newcommand{\tmp}[1]{\textcolor{blue}{#1}} 
\newcommand{\rr}[1]{\textcolor{black}{#1}} 

\newcommand\figref[1]{Fig.~\ref{#1}}
\newcommand\secref[1]{Sec.~\ref{#1}}
\newcommand\tabref[1]{Table~\ref{#1}}
\newcommand\appref[1]{\ref{#1}}

\journal{}
\begin{document}

\begin{frontmatter}



\title{Balancing the Norwegian regulated power market anno 2016 to 2022}


\author[inst1]{Pål Forr Austnes}
\affiliation[inst1]{organization={SINTEF Energy Research},
            addressline={Sem Sælands vei 11}, 
            city={Trondheim},
            postcode={7034}, 
            country={Norway}}

\author[inst2]{Signe Riemer-Sørensen}
\affiliation[inst2]{organization={SINTEF Digital},
            addressline={Pb. 124 Blindern}, 
            city={Oslo},
            postcode={0314}, 
            country={Norway}}

\author[inst3]{David Andreas Bordvik}
\affiliation[inst3]{organization={Akershus Energi},
            addressline={Brogata 7}, 
            city={Lillestrøm},
            postcode={20000}, 
            country={Norway}}

\author[inst1]{Christian Andre Andresen}

\begin{abstract}
The balancing market for power is designed to account for the difference between predicted \rr{supply/demand of electricity and the realised supply/demand}. However, increased electrification of society changes the consumption patterns, and increased production from renewable sources leads to larger un-predicted fluctuations in production, both effects potentially leading to increased \rr{balancing}. We analyse \rr{public} market data for the balancing market (manual Frequency Restoration Reserve) for Norway from 2016 to 2022 to investigate and document these effects. The data is newer than for \rr{similar} analyses and the eight years of data is more than double the time span \rr{previously covered}.

The main findings are: a) The balancing volumes are dominated by hours of zero regulation but \rr{for non-zero hours}, the balancing volumes are increasing during the eight-year period. b) The balancing prices are primarily correlated with day-ahead prices and secondary with balancing volumes. The latter correlation is found to be increasingly non-linear with time. c) The balancing volumes and the price difference between balancing price and day-ahead price are strongly correlated with the previous hour. d) The \rr{increasing} share of wind power has not impacted the frequency of balancing, which has remained stable during the 8 years studied. However, the volumes and share of balancing power compared to overall production have increased, suggesting that the hours which are inherently difficult to predict remain the same. e) Market data alone cannot predict balancing volumes. If attempting, the auto-correlation becomes the main source of information.
\end{abstract}




\end{frontmatter}



\include{text}




\bibliographystyle{elsarticle-harv} 
\bibliography{bibliography}





\end{document}

%% file: text.tex
\section{Introduction}
\subsection{Motivation}
The Nordic power market is a so-called deregulated market where the balancing market is a tool for the Transmission System Operator (TSO) to compensate the gap between the planned production and demand that has been settled in the day-ahead market and the actual production and demand. Electrification and an increased volume in the electricity market combined with increasing renewable energy from, e.g., wind, results in more volatile production and enlarge the need for grid balancing. Hence, the deregulated electricity market becomes more challenging \rr{for the TSO to balance, but also for the power producers who need to become more flexible and provide more capacity for balancing.}
The electricity volumes for all producers and consumers are settled 12--36 hours in advance of the operational hour in the day-ahead market. However, neither consumption nor production can be predicted with 100\% certainty. The producers can either balance their position through trade in a continuous intra-day market (such as XBID\footnote{\url{https://www.epexspot.com/}}) or settle them in the balancing market \citep{skytte_regulating_1999}. 
Deeper understanding of the balancing market may provide the power producers with decision support for improved production planning and strategic decisions regarding which market to settle potential imbalances in. 
\rr{In this work we provide a thorough analysis of public data for the Norwegian deregulated power market, which is coupled to the Nordic markets. In addition to providing insights about the Norwegian market, it serves as example for other deregulated markets and we provide the code for others to repeat the analysis\footnote{\url{https://github.com/SINTEF/balancing-market-analysis}}. Insight into and understanding of the statistical properties for this market is crucial for anyone endeavouring to model it, and is a natural part of \textit{exploratory data analysis}, hence this paper paves the way for others to better understand these markets and carry the analysis over to other markets.}

\subsection{The Nordic power market}
\begin{figure}[!ht]
\centering
\includegraphics[width=0.9\linewidth]{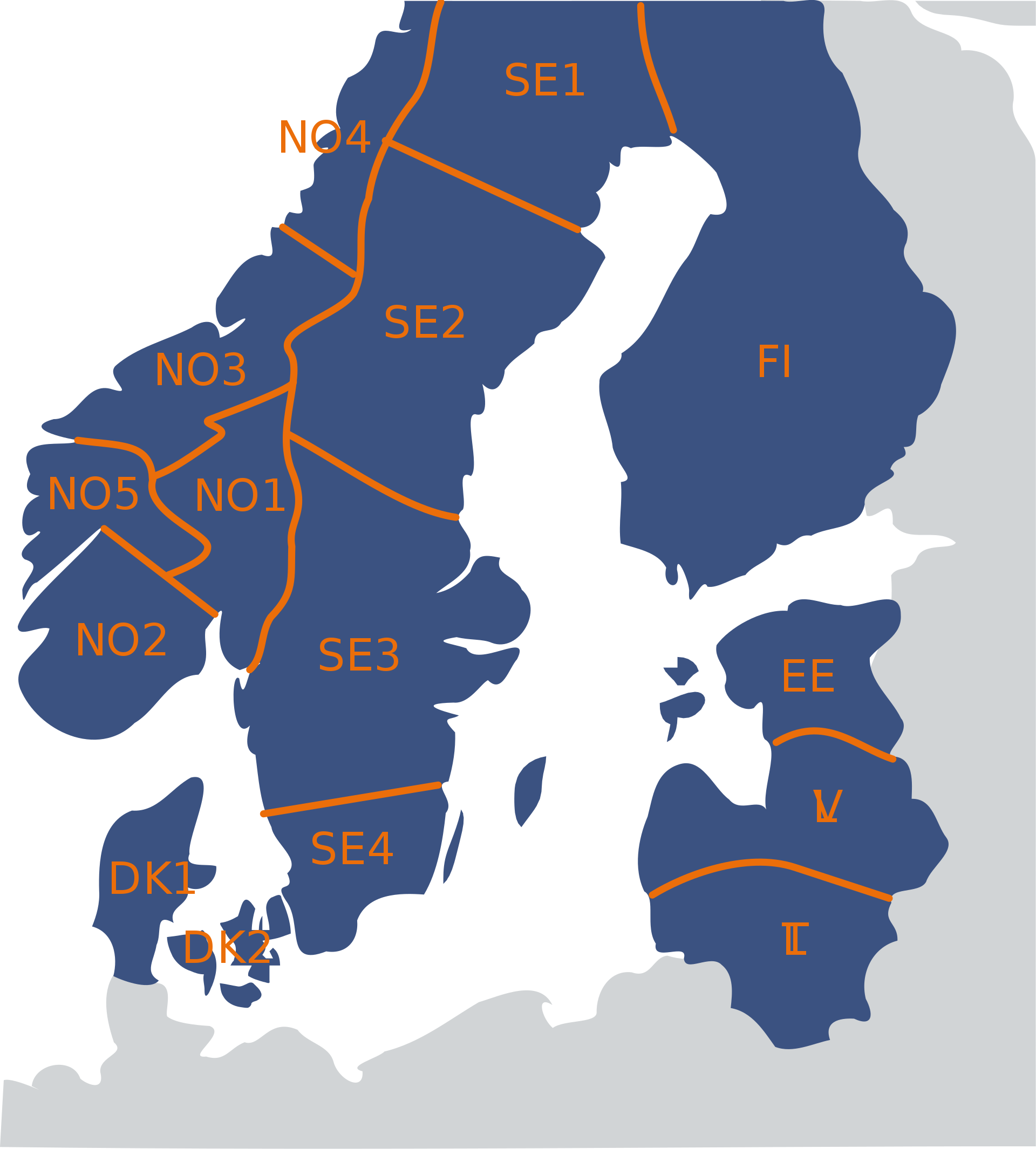}
\caption{The Norwegian power market and neighbouring zones throughout the period 2016-2022. The \rr{Norwegian zones NO1-5} are in part defined by bottlenecks in the transmission grid. \rr{The Swedish (SE1--4), Danish (DK1--2), Finnish (FI), and Baltic (EE, LV, LT) zones are regulated by the respective TSOs.} Figure from \citet{FigElectricityPriceArea}.}
\label{fig:zones}
\end{figure}

The Nordic power sector consists of 12 connected market zones as shown in \figref{fig:zones}, \rr{of which the five Norwegian zones (NO1--5) are under the TSO responsibility of Statnett. Zones in the neighboring countries are under the responsibility of the TSO of the respective country.} The zones are designed to take congestion in the transmission grid into account during market clearing. The electricity price within each zone is determined by the balance between demand, supply and exchange with neighbouring zones while respecting the transmission constraints \citep{nordpoolgroup_system_nodate}. Each zone is a unit in several markets; day-ahead (spot), intra-day, and the balancing market. 
The intra-day market allows market participants to settle imbalances directly between each other and thus avoid penalties from the TSO if their production or consumption differs from their day-ahead commitments. 

The TSO continuously monitors the imbalances of the \rr{transmission} grid and uses three principal mechanisms to restore the balance. Primary and secondary reserves are automatically activated to handle short and medium-term imbalances and are traded \rr{on D-1} and D-2 markets. If larger imbalances are detected or expected, the TSO can activate bids in the balancing market. Bids are either for up-regulation or down-regulation \rr{such that the TSO can increase or decrease production, respectively. In the following the term \textit{balancing} or \textit{grid balancing} refers to this action by the TSO.}
Up-regulation means that the producers are increasing their production to meet the needs of the TSO, which is policing the balancing by using the accessed reserves volumes to balance the grid. Alternatively one or more consumers may reduce their consumption, \rr{but historically this mechanism is of little importance, as the volumes have been small. This may change in the future}. In a similar manner down-regulation means decreased production or increased consumption. 
 Note that the regulation may or may not happen in the zone where the imbalance occurs (depending on available volumes and transmission capacity). 
 
The formation of the day-ahead prices and volumes \rr{is} a vast study topic. We do not dive into the underlying dynamics of the day-ahead market, but merely regard the day-ahead volumes and prices as given features when considering balancing volumes and prices.

\subsection{Outline}
In this paper we perform a statistical analysis of the market data for the Norwegian power market \rr{for the period 2016--2022} to document and describe changes. \rr{The aim is to enable future strategic} decision support \rr{primarily for power producers} based on public available data (in contrast to TSO-restricted sensitive data).
The paper is organised as follows: In \secref{sec:related_work} we recap related literature\rr{, and in \secref{sec:novelty} we} highlight our contributions. The data \rr{and sources of production} are presented in \secref{sec:data} \rr{together with some} general methodologies. We split the analysis of the balancing market into \secref{sec:volumes} focusing on volumes and \secref{sec:prices} focusing on prices. In particular we investigate the price premium defined as the difference between balancing price and day-ahead price. \secref{sec:relations} explores relations in the balancing markets before we attempt to identify the driving features through a simple model in \secref{sec:model}. Finally, we provide a brief summary and discussion in \secref{sec:results}.

\section{Related literature} \label{sec:related_work}
Research on analysing and modelling of the balancing volume in the Norwegian and Nordic markets is scarce, with only a handful of papers found \citep{skytte_regulating_1999, jaehnert_modelling_2009, klaeboe_benchmarking_2015, 10161961}. 
Research on the day-ahead and partially the intra-day markets have received more attention \citep{klaeboe_benchmarking_2015, hirth_balancing_2015, dimoulkas_forecasting_2016, hameed_applications_2021}. \citet{lago_forecasting_2021} presents a review of commonly used methods and benchmarks for day-ahead electricity price forecasting.

\subsection{The Nordic balancing market}
\citet{skytte_regulating_1999} investigated patterns in the balancing power market from an economical and business perspective on data from 1996. They established a relationship (correlation) between the day-ahead price and the balancing market price and linked it to the balancing volume by a linear model. \citet{jaehnert_modelling_2009} built upon \citet{skytte_regulating_1999} and proposed a linear statistical model (SARIMA) for the balancing volume. They modelled the effect of balancing volumes on electricity prices and social welfare using data from NO1 in 2007. 
\rr{A high linear correlation was identified} between balancing volume and the differences between day-ahead prices and balancing prices. The zones were significantly redefined in 2009 and 2010 with minor adjustments in 2013 \citep{zones}, which impact comparisons between \citet{jaehnert_modelling_2009} and later studies.

\citet{klaeboe_benchmarking_2015} is the most extensive research found covering both balancing volume and price forecasting in the Norwegian balancing market. The research benchmarked several time series-based forecasting models for balancing price and volumes in the balancing market such as different types of autoregressive models, Markov models, and arrival rate models for predicting the balancing state. The study considered market data from July 2010 to December 2012 (balancing volumes, overall production volumes, balancing states, balancing prices, and day-ahead prices). Good bidding strategies rely on forecasts, and for the producers it influences capacity allocations between markets. The importance of the balancing markets is increasing with the increased fraction of power from volatile renewable sources \citep{holttinen_estimating_2008}. \citet{klaeboe_benchmarking_2015} found that the correlation of \citet{jaehnert_modelling_2009} between price differences and balancing volume had declined from 0.7811 on 2007 data down to 0.47 on the 2010--2012 data. The aim of \citet{klaeboe_benchmarking_2015} was to predict the balancing market before the closing of the day-ahead market. However, they conclude that the balancing market is designed to handle unforeseen events and that leads to randomly distributed balancing volumes and prices. Nonetheless, an increasing fraction of volatile renewable energy in the market will increase the need for balancing and hence for improved model predictions both for the day-ahead market and for shorter timescales for the balancing market. 

In \citep{KLYVE2023121696}, bidding strategies between internal balancing and participation in the regulation market is investigated. The study concludes that in the current regulatory setting, profit maximisation for participants occurs when they allocate all available balancing resources to the balancing market and not to handle internal imbalances. The authors also highlight that knowledge of the market regulation price or the expected imbalances \rr{(i.e. a forecast)} can change the optimal bidding strategy.

\citet{dimoulkas_forecasting_2016} used a Hidden Markov Model on the Swedish bidding zone SE2 on 2014 market data to forecast both balancing price and volume. They investigated the partial auto-correlation for balancing volume and found the first and second lag significant, as well as significant correlation between balancing volumes and price differences. They achieved good predictions for one hour ahead but struggled to predict further, which they blamed on data being dominated by randomness. They did not utilise wind, solar, power production, and consumption load but proposed them for future work. 

In \citet{salem_2019}, the authors introduced a prediction methodology for intra-hour imbalance prediction from a TSO perspective, utilising data only available for the TSO. They applied quantile regression forest, which is an ensemble learning method with the ability to generate prediction intervals. The \rr{considered} features were market prices, planned transmission flows, production plans, and historical imbalances (lags), all with 5 minutes resolution for Norway (NO1--NO5) spanning January 2015 to December 2016. Some weather features were tried and found not to improve the result and hence excluded. However, the \rr{authors hypothesize} that weather may be influential for models incorporating consumption forecasts together with temperature forecasts or if the power \rr{systems} had a greater share of renewable wind or solar production.  

\rr{\citet{klaeboe_day-ahead_2022} studied bidding strategies between day-ahead and balancing markets for hydro-power producers in Norway. They concluded that coordinated bidding between the two markets is currently not profitable, but might be with increased volumes and price premiums. However, their model did not consider forecasting balancing volumes. In a recent study \citep{10161961}, a forecasting model of regulation volumes and price premiums based on Long Short Term Memory Neural Networks (LSTM) was developed for the five Norwegian market-areas. The model outperformed a simple baseline, but not to a significant degree. The authors underlined the lack of predictive power in the publicly available data. There are also commercial actors that have developed their own methodology and are offering prediction tools and services based at least partly on data-driven methods \citep[e.g.]{optimeering}.}

\subsection{Other markets}
\rr{In recent years, there has been a growing body of research on balancing volumes outside the Nordic region in Europe. The differences in market design, TSO practices and local conditions might limit the direct comparison between different studies, but the general methodologies and analyses are valid across different markets.}

\citet{garcia_forecasting_2006} applied artificial neural network techniques to forecast the balancing volume using data for England and Wales from 2001--2004. They explored various methods and achieved better performance for neural networks than conventional forecasting methods which was explained \rr{by} the neural nets ability to capture non-linear relationships between variables and influencing contributors. 

\citet{hirth_balancing_2015} investigated variable renewable energy sources, such as wind and solar production, in Germany, and the impact of forecast errors on balancing reserve requirements including the supply of balancing services by \rr{dispatchable} generators, and the incentives to improve forecasting provided by imbalance charges. The essence of their finding is that the balancing reserves depend on many factors and that wind and solar power forecast errors are two of several possible drivers, but that other factors are possibly more important in the balancing market. 

\citet{bottieau_2020} developed a quantile regression tool producing a probabilistic prediction of the future system imbalance. Furthermore, the tool quantifies risks and optimises participation of a market player in the imbalance market. The tool utilises an encoder-decoder neural network approach and was trained on historic imbalances along with historic production split on different sources from Belgium 2014--2018. Similar to \citet{salem_2019}, this data is not publicly available, but obtained via the TSO.

\rr{\citet{toubeau2022} developed a multi-horizon probabilistic forecasting tool for imbalance-volumes. Their model employs an attention-mechanism that highlights the influence from the model variables and therefore enhances explainability. \rr{Their model is trained} on Belgian TSO data from 2015--2020. \rr{The} model shows that for short-term forecasts, the auto-correlation is the most important variable. As the prediction horizon grows, cross-zonal transmission volumes give valuable information.}

\rr{\citet{engproc2023039018} presents a case study of the Greek balancing market and several short-term forecasting tools are developed and compared. The forecast-horizons considered are only 15-minutes and 1-hour and the authors does not claim their model has predictive power beyond this horizon. They state that additional explanatory variables should be considered, such as strategic bidding, operational, commercial and transmission constraints, as well as outages modelling. }

\rr{\citet{MERTEN2020114978} performed a thorough study and description of balancing markets in Europe, with an emphasis on Germany. They formulate different prediction models for price and deployment duration that are applicable both to secondary and tertiary reserves. These models are then used to formulate bidding strategies for market participants.}

\subsection{Data driven modelling of the Nordic balancing market}
\rr{The balancing market literature indicate challenges in providing long term (3-12 hours) robust predictions despite several attempts. While data driven approaches have been explored \citep[e.g.~][]{garcia_forecasting_2006, dimoulkas_forecasting_2016, KLYVE2023121696}, they can be difficult to tune.}

\rr{
The XGBoost package is a popular gradient-boost algorithm which has been used for both snapshot and time series prediction in the past \citep{Chen:2016:XST:2939672.2939785}. One of its major strengths is high out-of-the box performance on tabular data \citep{2021arXiv210603253S}. It has been shown to be efficient for industrial consumer demand in China and Ireland \citep{wang_short-term_2021}, and day-ahead forecasting based on smart-meters \citep{semmelmann_load_2022}. For wind power, the short term production has been addressed through similarity modelling based on to historic wind patterns using XGBoost \citep{zheng_xgboost_2019}. However, we are not aware of any applications to balancing market data, and hence we explore the framework as a method to investigate non-linear relationships that may remain non-significant in the statistical analysis.}

\subsection{Novelty and contribution} \label{sec:novelty}
We analyse \rr{public} market data for the Nordic balancing market from 2016 to 2022. The data is newer than for previous publications and the eight years of data span a longer time period than the three years previously analysed by \citet{klaeboe_benchmarking_2015} allowing us to consider changes over longer time spans. The main findings are:
\begin{itemize}
    \item In all zones, the fraction of intermittent renewable production has increased from 2016 to 2022.
    \item The balancing volumes are dominated by hours of zero regulation, despite increase in average balancing volume. The number of regulated hours is relatively stable, so the increased variation is due to larger balancing volumes.
    \item The partial auto-correlation of balancing volumes and of the price difference between balancing price and day-ahead price are strongly dominated by the previous hour.
    \item The balancing prices are primarily correlated with day-ahead prices and secondary with balancing volumes.
    \item The linear correlation between price premium and balancing volume was found to decrease from 2007 \citep{jaehnert_modelling_2009} to 2010--2012 \citep{klaeboe_benchmarking_2015}. We find a persistent correlation but shifted towards a more non-linear relationship.
    \item \rr{Through simple modelling, we show that} market data alone cannot predict balancing volumes or prices. We did not expect market data alone to hold sufficient predictive power, which was confirmed by the study. If attempting, the auto-correlation becomes the main source of information limiting the time horizon of the predictions.
\end{itemize}


\section{Data and methodology} \label{sec:data}
\subsection{Data}
All data is public historical data provided by Nord Pool\footnote{System price and Area price calculations, \url{https://www.nordpoolgroup.com/trading/Day-ahead-trading/726
Price-calculation/}} and Entso-E\footnote{Entso-E python module \url{https://github.com/EnergieID/entsoe-py.667
original-date: 2017-07-12T13:17:39Z}}. 
The numerical market data covers the period from 2016-01-01 to 2022-11-30 resulting in 60600 hours sequenced time-series data in total. Occasionally missing values are imputed with the value of the previous hour. \rr{The investigated variables are listed in \tabref{tab:variables}.}

\begin{table}
\centering
\begin{tabular}{l | l} 
\textbf{Source} & \textbf{Variable} \\
 \hline
 \hline
Entso-E & Production Hydro Water Reservoir \\
Entso-E & Production Hydro Run-of-river and poundage \\
Entso-E & Production Wind Onshore \\
Entso-E & Production Hydro Pumped Storage \\
Entso-E & Production Fossil Gas \\
Entso-E & Production Sum other \\
        & (Biomass, Waste, Other renewable, Other) \\
Nordpool & Capacity for down regulation \\
Nordpool & Capacity for up regulation \\
Nordpool & Consumption volume  \\
Nordpool & Day ahead buy volume \\
Nordpool & Day ahead sell volume \\
Nordpool & Day ahead price \\
Nordpool & Realised production volume \\
Nordpool & Realised balancing volume \\
Nordpool & Balancing up prices \\
Nordpool & Balancing down prices \\
\end{tabular}
\caption{List of the investigated variables. Entso-E data is used only for the Norwegian zones (NO1--5) while some Nordpool variables are also investigated for the neighbouring zones (NO1--5 plus DK2, FI, SE1--3).}
\label{tab:variables}
\end{table}

\begin{figure*}[t]
    \centering
    \includegraphics[width=\textwidth]{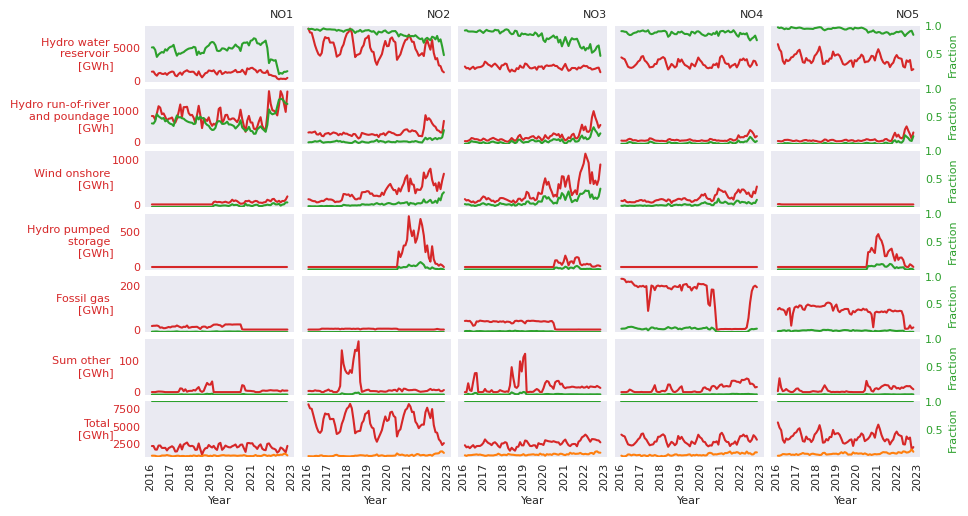}
    \caption{The power production composition in the Norwegian market zones (columns) shown as monthly averages (red lines, left axis) and fractions of total for the same month and zone scaled between 0 and 1 (green lines, right axis). The bottom row shows the total production in each zone and the fraction of electricity production used for balancing within the same month (orange, right axis). Historically, hydropower has dominated, but for all zones there is a relative decrease in hydropower since 2016 (top row, green lines), and an increase in volatile sources e.g. onshore wind (third row, green lines). On a monthly basis, the balancing only makes up a few percent of the total production volume (bottom row, orange line). NO2 is the zone with larges absolute production and largest seasonal variations (bottom row, second column, red line). NO3 has the largest relative fraction of directly weather based production from from run-of-river and onshore wind.}
    \label{fig:production_sources}
\end{figure*}

\rr{\figref{fig:production_sources} shows the monthly averages of different electricity sources in each of the Norwegian zones. Historically, hydropower has dominated the Norwegian power production but new sources emerge and change the mix. For all zones there is a relative reduction in the fraction of reservoir hydro power from 2016 to 2022 and an increase in volatile weather-based sources such as wind and run-of-river hydropower. NO2 is the zone with the largest absolute energy production in Norway, and also the one with largest seasonal variations. NO3 has lower overall production, but has the largest fraction of directly weather-based production from run-of-river and onshore wind. In general, the balancing is only a small, but increasing, fraction of the total volume. NO5 has the highest balancing fraction, but due to larger overall volume, NO2 has the highest balancing volume in absolute numbers. This is due to a large fraction of flexible hydro reservoir power in the zones, in combination with high transmission capacity to the population dense zones NO1--2.}

\subsection{Methodology} \label{sec:methodology}

\rr{For each zone, we investigate the statistical properties over time. We consider the daily and monthly patterns of production and consumption as well as dominating time scales through discrete Fourier transforms. We inspect the distributions of up and down regulation and balancing prices in total, and the temporal evolution through the six-month moving average.}

\rr{\citet{klaeboe_benchmarking_2015} concluded that the balancing volumes and prices are randomly distributed because the market is designed to handle unforeseen events. Random data can be characterised as white noise, meaning one cannot foresee the value of future time steps given historical data. }

Hence we investigate the time development of mean, variance, correlation and stationarity of regulation volume and balancing prices.


\rr{The auto-correlation measures the correlation between the signal itself and a lagged copy of itself. It quantifies if past data is related to current data, which is particularly important for univariate models which generate forecasts based on a single time series. The auto-correlation contains contributions from both indirect and direct correlations between the current time and prior time steps \citep{brownlee_gentle_2017}. Consequently, the correlation at lag 2 can be significantly influenced by, and dependent on, the correlation for lag 1. The partial auto-correlation corrects for this, and measures the correlation that remains after removal of correlations from shorter lags \citep{cowpertwait_correlation_2009}.}

\rr{Statistical frameworks for time series modelling such as auto-regressive integrated moving average (ARIMA) requires the time series to be stationary or that it can be made stationary by removal of trends and seasonal variations \citep{box2015time}. A stationary time series is described by a stochastic process for which the unconditional joint probability distribution remains unchanged when shifted in time. 
The most intuitive interpretation of stationarity is that the statistical properties are allowed to change over time but only if the way they change does not itself change over time. With predictive modelling in mind we investigate temporal evolution of mean and standard deviation of the balancing market features.}

\rr{In addition we compute the Augmented Dickey-Fuller statistic and p-value for selected time series\footnote{Using the python package statsmodels version 0.13.5}. For a threshold p-value of $0.01$ the presence of a unit-root indicating non-stationarity can be rejected/confirmed at 99\% confidence level if the p-value is greater/smaller than the threshold.}

\rr{For exploring relations between available variables, we apply Spearman rank correlation as it handles outliers and non-linear relations better than the linear Pearson correlation coefficient \citep{auffarth_machine_2021}.}

\rr{Finally, we make an attempt at a simple XGBoost model of balancing volumes to check for any non-linear relations and combinations of factors that may have been overseen by the statistical analysis \citep{Chen:2016:XST:2939672.2939785} and explore the feature importance using Shapley Additive Global importancE (SAGE) values \citep{NEURIPS2020_c7bf0b7c}. SAGE estimates how much each feature contributes to the model's predictive power.}

\section{\rr{Production, consumption and balancing volumes}} \label{sec:volumes}
\subsection{General observations in the day-ahead market}
\begin{figure*}
    \centering
    \includegraphics[width=0.9\linewidth]{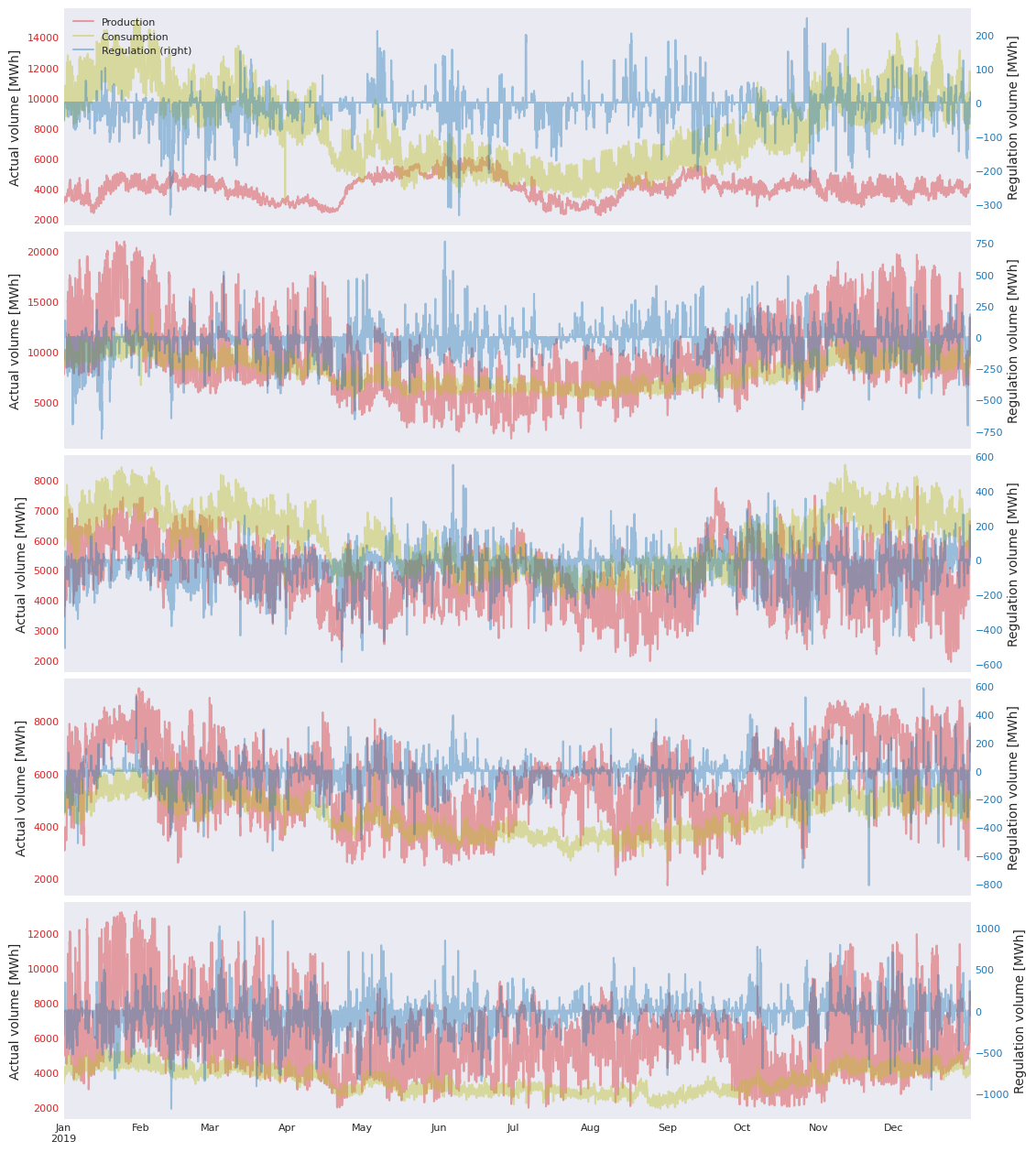}
    \caption{Hourly actual production (red, left axis), consumption (yellow, left axis) and balancing volumes (blue, right axis) for 2019 for each of the Norwegian zones from NO1 (top panel) to NO5 (bottom panel). \rr{In all zones, we observe that consumption is higher in summer than in winter (yellow). Some zones are roughly balanced between consumption and production (e.g. NO2) while others are dominated by consumption (e.g. NO1) or production (e.g. NO5). This will naturally affect the regulation volumes as well (blue). Since regulation is a correction to the day-ahead plan, it can take on both positive and negative values, and does not follow any obvious pattern.}}
    \label{fig:market_vol}
\end{figure*}

The day-ahead markets constitute the majority of available power traded and while focusing on the balancing market, we find it useful to also include some analysis of the day-ahead market for better understanding of the balancing market. 
\figref{fig:market_vol} shows the hourly production, consumption and regulation for the Norwegian zones in 2019. We observe \rr{a} clear seasonal pattern in the production and consumption volumes. This would be similar for other years before 2020 \rr{when} the covid-19 pandemic significantly changed consumption and 2022 \rr{when} the geopolitical situation around Ukraine affected energy-trade in Europe.  

Some zones are roughly balanced between consumption and production (e.g. NO2) while others are dominated by consumption (e.g. NO1) or production (e.g. NO5). \rr{Hence production does not necessarily happen in \rr{the} same geographical region as consumption.} This will naturally affect the regulation as well, \rr{which may not happen in the zone where the under- or over-production happens}. The first noticeable difference between balancing and production/consumption volumes is the fact that regulation is a correction to day-ahead production and can take on both positive and negative values. The balancing volumes are centered near zero, but in contrast to the day-ahead volumes, the balancing volumes do not reveal any systematic patterns and appear to be stochastic in nature.

\subsection{Temporal patterns in balancing volumes}
\begin{figure}[!tb]
\centering
    \includegraphics[width=0.9\linewidth]{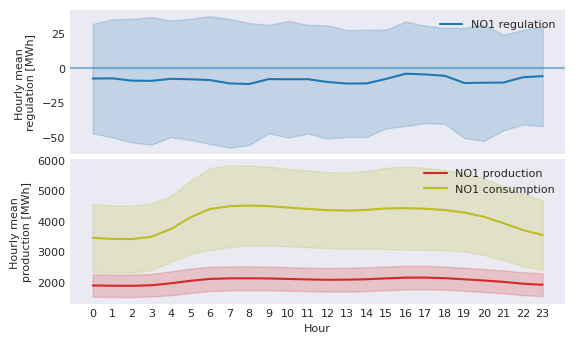}
    \includegraphics[width=0.9\linewidth]{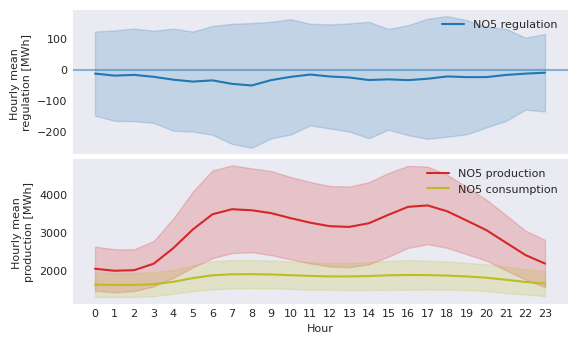}
    \includegraphics[width=0.9\linewidth]{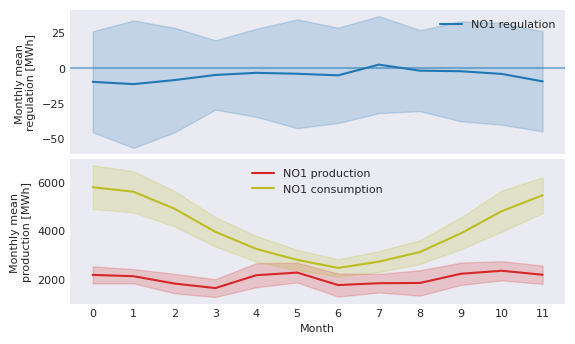}
    \includegraphics[width=0.9\linewidth]{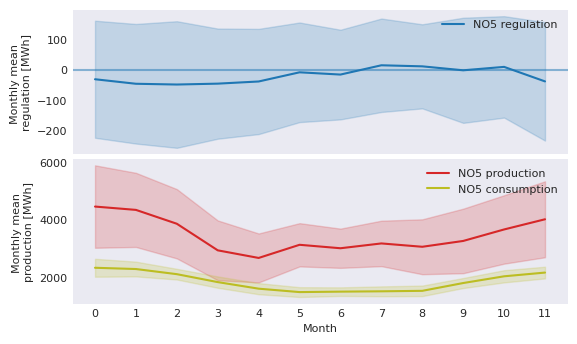}
    \caption{{\it Upper two panels}: Hourly averages and standard deviations (shaded) of 2019 data for production/consumption (red/yellow) volumes and corresponding balancing (blue) for NO1 and NO5. The characteristic daily pattern is visible in consumption and production, while there is no clear pattern in the balancing. {\it Lower two panels}: Monthly averages and standard deviations of 2016--2022 data for NO1 and NO5. The consumption and production show a characteristic seasonal pattern, while the balancing fluctuates over the entire year. In both cases, we show NO1 which is consumption-dominated and NO5 which is production-dominated, but similar patters occur for the other zones as well.}
    \label{fig:daily}
\end{figure}

The upper two panels of \figref{fig:daily} show the balancing, production and consumption volumes per hour of the day averaged over the year of 2019 for NO1 which is consumption-dominated and for NO5 which is production-dominated. The production and consumption follows a double-hump curve. During night, less power is consumed/produced before peaking around 06:00 when most people start their days. This is followed by a slight decrease during daytime to another peak around 18:00, matching evening household activities. Note that a large portion of major industrial consumers are not trading electricity in the day-ahead market and hence private consumption dominate the pattern. The regulations are dominated by a small negative mean, and stable standard deviation across the day. We observe no significant difference between weekdays, and neither between working days and holiday (not shown).

The two lower panels of \figref{fig:daily} show the balancing, production and consumption per month averaged over 2019--2022 for NO1 and for NO5. The production and consumption follow a seasonal pattern with increase in winter and decrease in summer. The balancing volumes are dominated by a small negative mean, and stable standard deviation across the year. 


\subsection{Balancing volume statistics}
\begin{figure*}[t]
    \centering
    \includegraphics[width=0.9\linewidth]{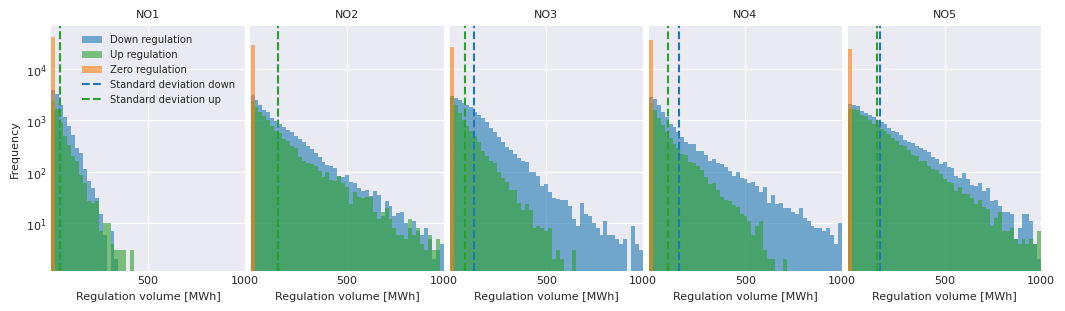}
    \includegraphics[width=0.9\linewidth]{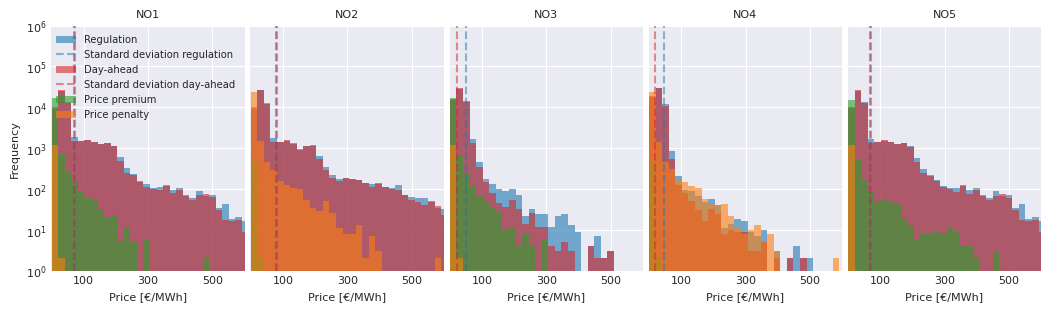}
    \caption{{\it Upper panel:} Balancing volume histograms for the period 2016--2022: Actual balancing volumes for frequency of hours with zero regulation (orange, first bin), non-zero up-regulation (green) and absolute values of non-zero down-regulation (blue). The vertical dashed lines shows the standard deviation of the down- and up-regulations. Similar behaviours are present if the balancing volumes are normalised to actual production. The evolution of regulation volumes and potential asymmetry over time is shown in \figref{fig:regulation_volume_moving_avg}. {\it Lower panel:} Histograms of day-ahead and balancing prices for 2016--2022 for each of the Norwegian zones. We observe that the distributions seem to consist of two overlapping distributions (note the logarithmic scale): A narrower Gaussian centred around the median price (40~€/MWh) and a wider distribution of higher prices. A few counts with prices higher than 700~€/MWh are outside the plotting range.}\label{fig:regulation_volume_histogram}
\end{figure*}

\rr{The upper panel in} \figref{fig:regulation_volume_histogram} shows the distributions of balancing volumes for each of the Norwegian zones. The corresponding statistics are given in \tabref{tab:global_stats} in \appref{app:Appendix}. 
Notice that the y-axis is logarithmic and hence the distributions are clearly dominated by hours of zero regulation. The high zero fraction is expected, \rr{since balancing is a deviation from the day-ahead plan per design}. 

We observe that \rr{for some zones (e.g. NO1),} the distributions are almost symmetric \rr{in up and down regulation}, while it is asymmetric for others (NO3--4). \rr{The distributions are heavier on the small regulations than a Gaussian, and are more similar to a Laplacian in shape.} Where asymmetric, \rr{it is always the down-regulation which is more frequent than up-regulation}. This is also reflected in \tabref{tab:global_stats} (in \appref{app:Appendix}) where all zones \rr{have} small negative mean values indicating a skewness towards down-regulation. This is expected as upwards regulation is limited by capacity within the zone, while downwards regulation more often is available in the hydropower dominated zones. 
    
By normalising the balancing volumes with the actual production of the specific hour, we can conclude that the distributions of balancing volumes and their skewness are not strongly related to the overall production volume in each zone. 

\subsection{Temporal development of balancing volumes} \label{sec:temporal_vol}
\begin{figure*}
    \centering
    \includegraphics[width=0.9\linewidth]{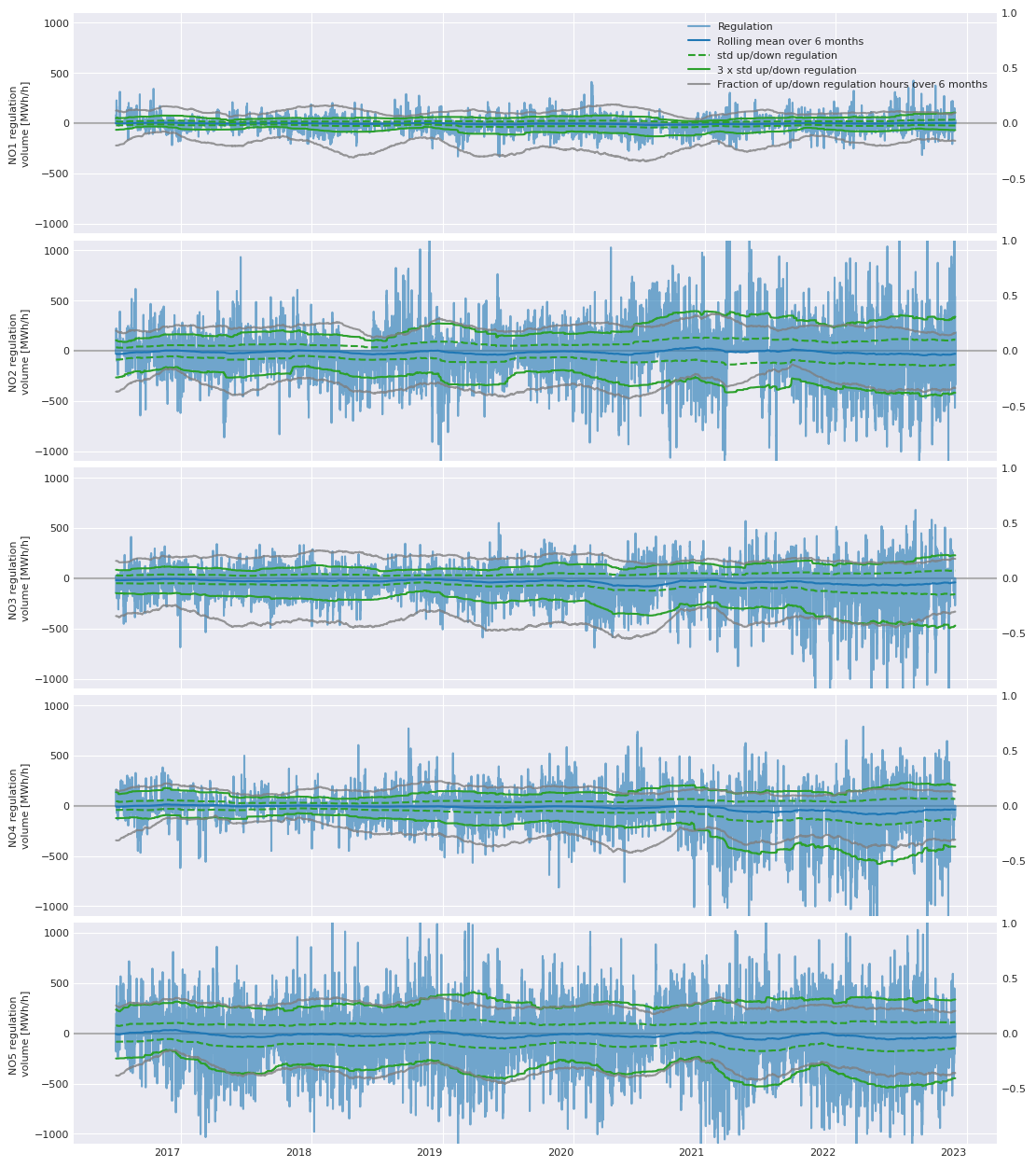}
    \caption{Balancing volumes (faint blue) and six-months moving average for one hour incremental rolling from 2016 to November 2022 (blue) for NO1 (top) to NO5 (bottom). The dashed green line shows the standard deviation and solid green is three standard deviations. The grey line indicates number of hours with non-zero regulation for a rolling window of six months. \rr{The mean and one standard deviation appear constant over time (blue and dashed green lines), but the $3\sigma$ lines (green solid) reveal a trend towards increasing variation in regulation for some zones. This trend remains when normalising to the actual production in each zone.} 
    }
    \label{fig:regulation_volume_moving_avg}
\end{figure*}

\rr{To investigate whether the balancing is random or systematically changing}, \figref{fig:regulation_volume_moving_avg} shows the balancing volumes and the six-month moving average for NO1--5. The mean and one standard deviation appear constant over time, but the $3\sigma$ lines reveal a trend towards increasing variation in regulation for some zones. This trend remains when normalising to the actual production in each zone. In addition we have checked for unit roots using the Augmented Dickey-Fuller test for day-ahead, production, consumption and balancing volumes. At the 99\% confidence level there are no overall constant, linear or quadratic trends in the hourly volumes. From this we cannot conclude the presence of any non-stationary evolution, but given the trend in the variance, we encourage caution if applying model frameworks that rely on stationary data. The partial auto-correlation indicates that only the correlation at lag $h-1$ significantly describes the present time step. 

The standard deviation of the down-regulation increases more than the up-regulation. This may be caused by an increasing number of hours of with overproduction compared to planned production from volatile electricity sources, such as wind. \rr{This is particularly pronounced in NO3--4, where wind production have increased the later years}.

\figref{fig:regulation_volume_moving_avg} also shows the fraction of hours with non-zero balancing volumes for a rolling window of six months. There is some variation, but no systematic trend over time. This indicates that the driver for increase in balancing volumes is on the size of the volumes rather than the number of hours with balancing.

\begin{figure*}[tbp]
    \centering
    \includegraphics[width=0.9\linewidth]{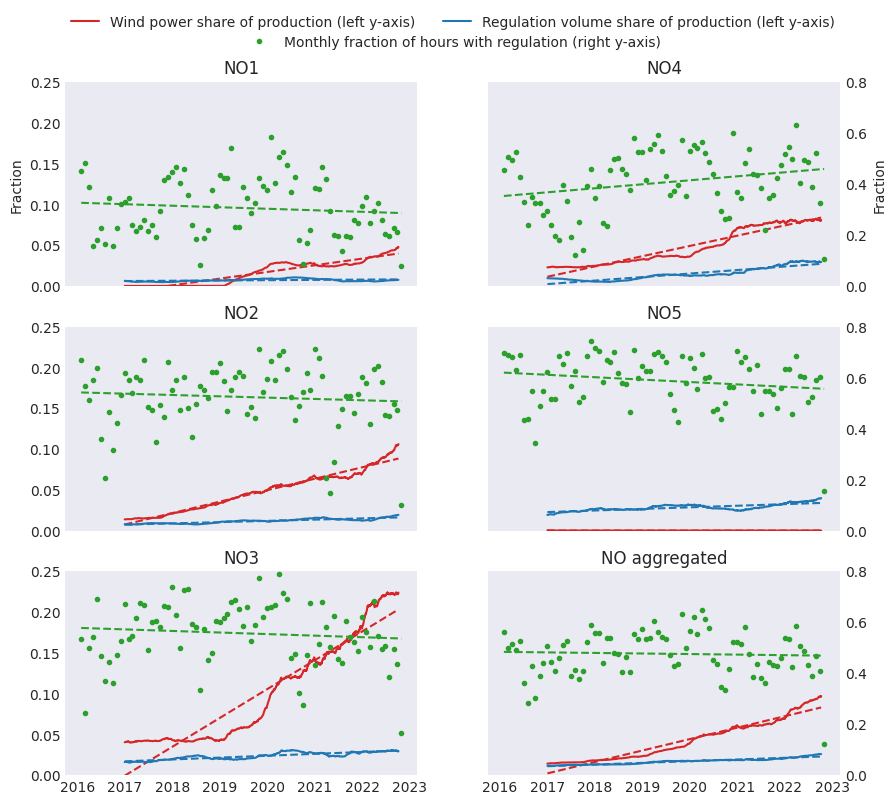}
    \caption{\rr{Comparison of increase in wind power production with need for balancing.} Left y-axis: Wind production share of total production (red solid lines) and absolute regulation volume share of total production (blue solid lines) in NO1--5 and aggregated for all zones. Right y-axis: Monthly share of hours with activated regulation reserves (green dots) in NO1--5 and the average of all zones aggregated. The dotted lines represent a linear best fit of the different variables. \rr{Apart from NO5, all zones see an increase in wind power production (red lines) which is larger than the corresponding increase in regulation (blue lines). The number of hours with balancing remains almost constant (green lines), indicating that the total increase in balancing volume is driven by larger volumes rather than more hours with need for balancing}.
    }
    \label{fig:ratio_evolution}
\end{figure*}

In \figref{fig:ratio_evolution} we show the interplay between the number of hours with activated reserves and fraction of wind power, and regulation volumes compared with the overall production in all Norwegian zones. During the eight years of analysed data, the production from wind has increased significantly in all zones except NO5. At the same time, we observe only a small increase in the share of regulation volume compared to the overall production. The fraction of hours with activated reserves show a mild negative trend in all zones, except NO4. This suggests that while volumes in the regulating market are increasing, the occurrence of events which require activation of regulating power does not increase. I.e. the number of hours where the day-ahead cleared volumes does not match the actual consumption are similar both with and without a large wind power share. Furthermore, it suggests that the large introduction of wind power does not cause more hours of activated reserves, rather the hours that were hard to predict historically remain hard to predict today.

\subsection{Periodicity of balancing volumes}
\begin{figure*}
    \centering
    \includegraphics[width=0.9\textwidth]{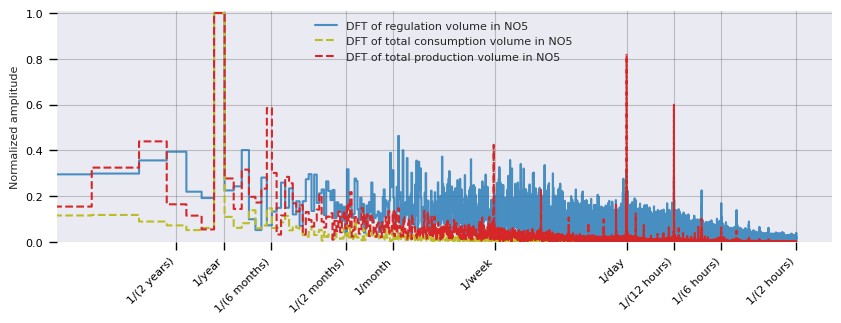}
    \caption{Discrete Fourier transform of balancing (blue), consumption (yellow) and production (red) volumes in NO5 for 2016-2022. The amplitudes are normalized to the largest observed frequency spike. NO1-4 show qualitiatively similar results. \rr{All zones have similar spikes with the dominant frequency being yearly, followed by daily (12 and 24 hours) coming from the two-hump pattern in \figref{fig:daily}}.}
    \label{fig:regulation_volume_DFT}
\end{figure*}

Since the periodic and seasonal patterns in the balancing volume are vague, we perform a further investigation of periodicity. A discrete Fourier transformation converts the data from time-domain to frequency-domain and individual frequencies may reveal periodic patterns. \figref{fig:regulation_volume_DFT} shows the discrete Fourier transform\footnote{Computed with Tensorflow.signal.rfft \url{https://www.tensorflow.org/tutorials/structured\_data/time\_series}} for both production, consumption, and balancing volumes for NO5 for the period 2016-2022. All zones have similar spikes with the dominant frequency being yearly, followed by daily (12 and 24 hours) coming from the two-hump pattern in \figref{fig:daily}.
Consumption volumes are the most periodical with little noise, while production volume periodicity is more variable and depends on the mix of production sources in the zone. The balancing volume periodicity also shows a major frequency at one-year intervals, but the rest of the spectrum is noisy with no clear periodicity.

\subsection{Balancing capacity}
As shown in \figref{fig:regulation_volume_histogram}, down-regulation is more frequent than up-regulation. This is naturally linked to available production capacity. \figref{fig:regulation_capacity_moving_avg} (in \appref{app:Appendix}) shows the production capacity declared available for regulation for each of the Norwegian zones along with six-month rolling averages. In general, the capacity for down-regulation is large when the capacity for up-regulation is small, but the sum is not constant as they also depend on regular production. The capacities available for balancing have clear seasonal variations. \rr{NO2 clearly dominates the regulation capacity, followed by NO5 as discussed in \secref{sec:data}.}

\section{Day-ahead and balancing prices} \label{sec:prices}
Balancing prices consist of two prices, one for up-regulation and one for down-regulation. The regulation prices are also defined for hours with no regulation. In the majority of hours, the balancing prices are identical to the day-ahead prices. However, when deviating the up-regulation price is higher than the day-ahead, while the down-regulation price is lower than day-ahead. If not specified, we consider the up-regulation price in the following, but qualitatively similar conclusions apply for the down-regulation prices.

\subsection{Temporal patterns in the prices}
\begin{figure}[htb]
    \centering
    \includegraphics[width=0.9\linewidth]{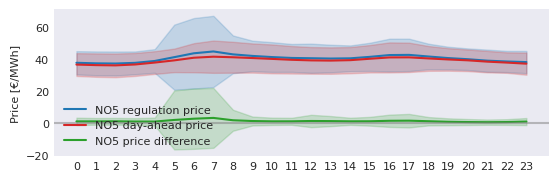}
    \includegraphics[width=0.9\linewidth]{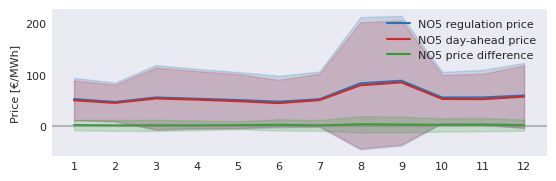}
    \caption{{\it Upper panel:} Hourly averages and standard deviations (shaded) of day-ahead price (red) and balancing price (blue), and the difference between the two (green) for 2019 data for NO5. {\it Lower panel:} Monthly averages and standard deviations (shaded) day-ahead price and balancing price for 2016--2022 for NO5. In both panels, we show NO5 which is production-dominated. The other regions and years show similar patterns for most years. A similar plot for weekdays do not reveal any significant difference between weekdays.}
    \label{fig:price_patterns}
\end{figure}

\begin{figure}[htb]
    \centering
    \includegraphics[width=0.9\linewidth]{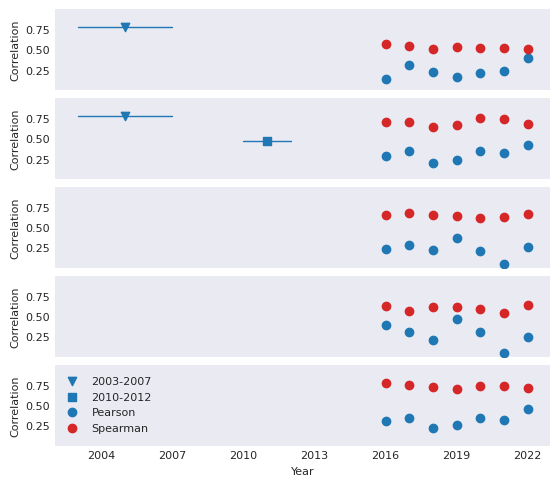}
    \caption{Pearson (blue) and Spearman (red) correlations between balancing volumes and price differences for each year for NO1 (top) to NO5 (bottom). We compare to previous values for 2003--2007 \citep{jaehnert_modelling_2009} and 2010--2012 \citep{klaeboe_benchmarking_2015}. The 2003--2007 value was based on NO1 data, before it was split into NO1 and NO2 and is thus shown for both zones.}
    \label{fig:delta_correlations}
\end{figure}

The upper panel in \figref{fig:price_patterns} shows the characteristic daily pattern for hourly averages and standard deviations of day-ahead price and balancing price for NO5 for 2019. There are hints of the two-hump shape also seen in the volumes in \figref{fig:daily}, but less pronounced. The balancing prices are strongly correlated with the day-ahead prices, however, in the morning, the variance is larger for balancing price than for day-ahead. This pattern is seen in all the Norwegian zones, and for several years. A possible explanation can be the uncertainty in the timing of the morning peak (rather than the magnitude) leading to larger variation in the balancing prices.

The lower panel in \figref{fig:price_patterns} shows the monthly average and standard deviation for the prices. We observe the same trends as for volume (lower panels in \figref{fig:daily}) where the prices follow the demand and are on average higher in winter than in summer. However, where the balancing volumes appear as random noise over the year, the balancing prices overall follow the day-ahead prices.

\subsection{Balancing price statistics}
\rr{The lower panels of} \figref{fig:regulation_volume_histogram} shows the histograms of day-ahead and balancing prices for 2016--2022 and the statistics are given in \tabref{tab:global_stats} in \appref{app:Appendix}. We observe that the distributions seem to consist of two overlapping distributions (note the logarithmic scale): A narrower Gaussian centred around the median price (40~€/MWh) and a wider distribution of higher prices. However, this multi-distribution shape may be due to temporal evolution of the prices.

\subsection{Temporal evolution of the balancing price} \label{sec:temporal_price}
\begin{figure*}[thbp]
    \centering
    \includegraphics[width=0.9\linewidth]{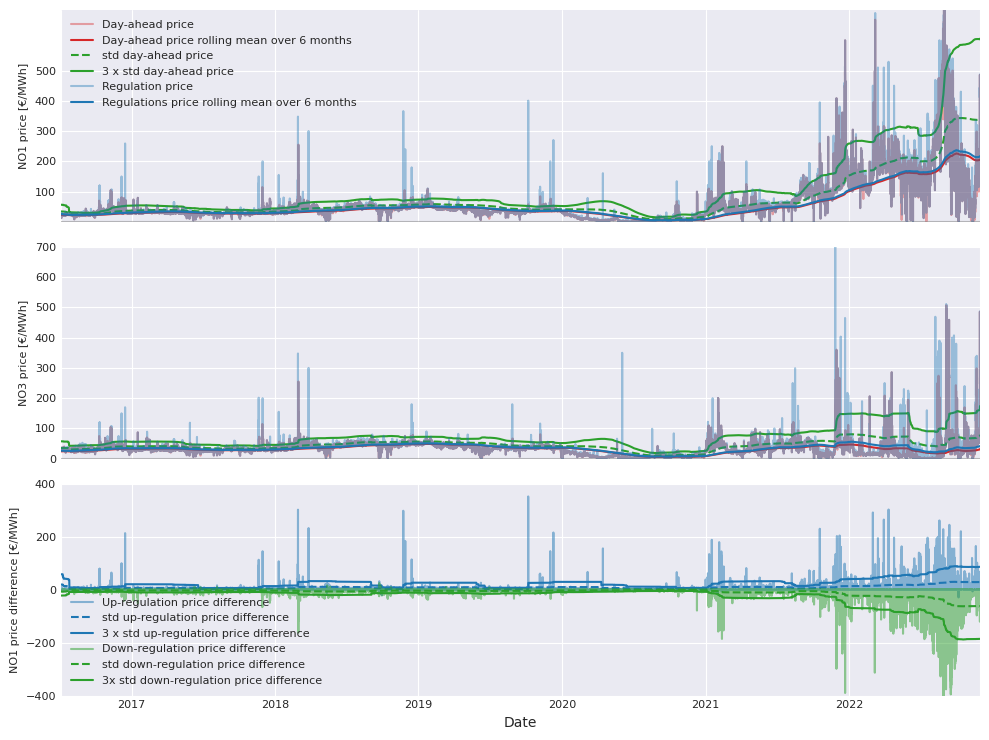}
    \caption{{\it Upper and middle panels:} Per hour day-ahead and balancing prices together with the six-month moving average and standard deviation for balancing price calculated using one hour incremental rolling (backwards in time) from 2016 to November 2022 for NO1 (top) and NO3 (middle). \rr{A few counts with prices higher than 700~€/MWh are outside the plotting range. NO2, NO4--5 are shown in \figref{fig:regulation_price_moving_avg_full} in \appref{app:Appendix}. Due to large transmission capacity between zones, NO2 and NO5 are similar to NO1, while NO4 is similar to NO3.}
    \rr{{\it Lower panel:} The price difference (balancing minus day-ahead price) and the absolute difference together with the rolling standard deviation of the difference separately for up- and down-regulation prices for NO1. The spikes before 2021 are correlated in time with hours with large regulation, but not all balancing volume spikes leads to a price spike. 
    The average of price differences follows the pattern of the balancing prices, being relatively stable up until 2020, and with significant increase from 2021 onward.}}
    \label{fig:regulation_price_moving_avg}
\end{figure*}

\figref{fig:regulation_price_moving_avg} shows the day-ahead and balancing prices for \rr{NO1 and NO3 for} 2016--2022 together with six-month moving average and standard deviation calculated using one hour incremental rolling. First we notice that the two sets of prices are highly correlated, with the major difference being higher spikes in the balancing prices. Before 2020 the mean and standard deviations are rather stable, with some variation in 2020, and a systematic increase from 2021 onward. This increase is thought to origin in a combination of multiple meteorological and geopolitical factors such as covid-19 pandemic, below average precipitation, the commissioning of both the NordLink cable between NO2 and Germany, and the North Sea Link between NO2 and UK, as well as Russian invasion of Ukraine with subsequent significant changes to power trading patterns in Europe. 
This demonstrates that the mean and standard deviations are not constant over time, and that market couplings are particularly important for price formation.

\rr{The lower panel of \figref{fig:regulation_price_moving_avg} shows the difference between day-ahead price and balancing price (the price premium)}. The average of price differences follows the pattern of the balancing prices, being relatively stable up until 2020, and with significant increase from 2021 onward. This behaviour is reflected in all zones, but NO3--4 are more affected by individual outliers.

\commentblock{
\begin{figure*}
    \centering
    \includegraphics[width=0.9\linewidth]{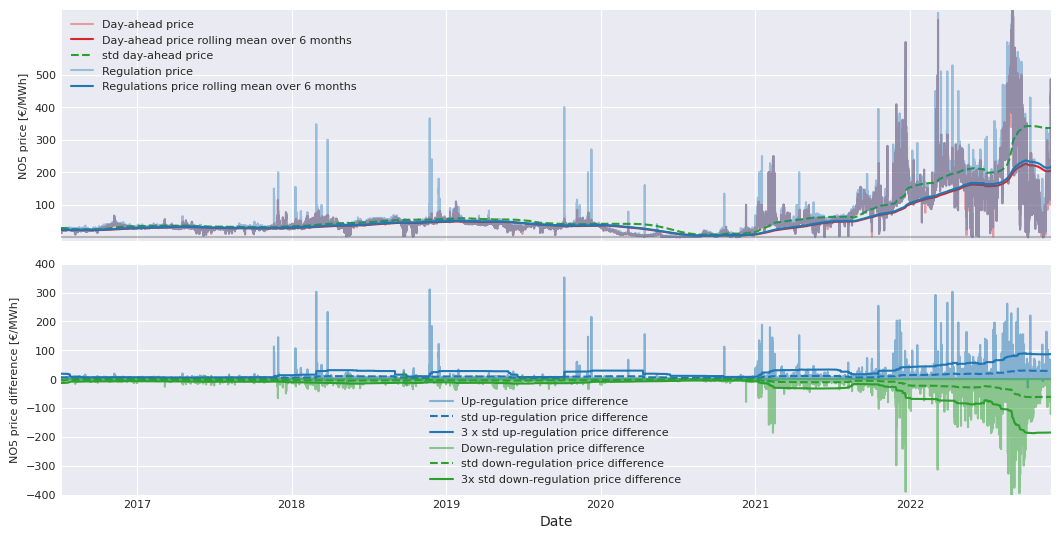}
    \caption{{\it Upper panel:} Per hour day-ahead and balancing prices for NO5 together with the six-month moving average and standard deviation for balancing price calculated using one hour incremental rolling (backwards in time) from 2016 to February 2022. {\it Lower panel:} The price difference (balancing minus day-ahead price) and the absolute difference together with the rolling standard deviation of the difference separately for up- and down-regulation prices. The spikes before 2021 are correlated in time with hours with large regulation, but not all balancing volume spikes leads to a price spike. 
    The average of price differences follows the pattern of the balancing prices, being relatively stable up until 2020, and with significant increase from 2021 onward. \rr{A few counts with prices higher than 700~€/MWh and down regulation price difference are outside the plotting range.}}
    \label{fig:regulation_price_moving_avg_NO5}
\end{figure*}}

The Augmented Dickey-Fuller test for unit-root in the day-ahead, balancing, and price premiums give no indication of unit roots for the period 2016--2022, and hence we cannot conclude non-stationarity beyond the variation in the data, but encourage caution if modelling using frameworks based on stationarity assumptions.

\subsection{Periodicity of prices}
\begin{figure*}
    \centering
    \includegraphics[width=0.9\textwidth]{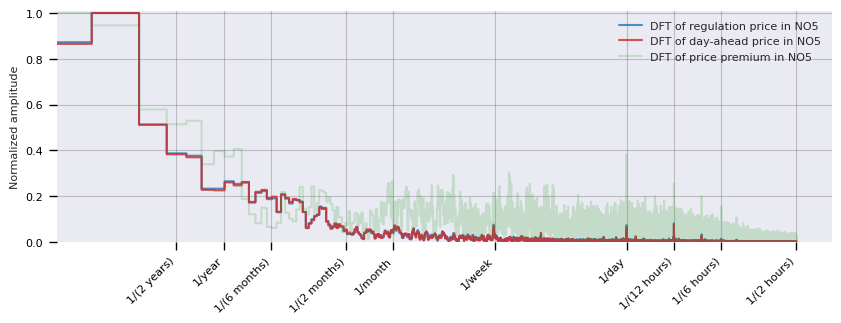}
    \caption{The discrete Fourier transforms of day-ahead price (red) , balancing price (blue, almost identical with day-ahead), and the difference between the two (green). The day-ahead and balancing prices show minor peaks at 12 and 24 hours, but they are partly washed out due to the overall temporal evolution. The price difference shows no particular pattern. \rr{All zones show similar lack of relevant time scales.}}
    \label{fig:DFT_price_NO5}
\end{figure*}

\figref{fig:DFT_price_NO5} shows the Fourier transforms of the prices and price differences. They do not reveal any systematic patterns, and we hypothesise that any seasonal patterns are masked by the overall temporal evolution. The partial auto-correlation indicates that only the first lag is significantly correlated with the original series, i.e. the temporal dependence is very short.

\section{Relations in the market} \label{sec:relations}
In this section we investigate correlations between various market factors.

\subsection{Market relations for volumes}
\figref{fig:correlation_allzone_volumes} \rr{in \appref{app:Appendix}} shows the Spearman correlation between balancing volumes of the Nordic market zones plus neighbouring zones (green box), consumption and production volumes (black box, orange box in \figref{fig:correlation_allzone_volumes}), volumes from various production types and available balancing capacity. First, we observe that the balancing is correlated between zones, in particular NO1--2--5 and NO3--4 but also with neighbouring zones outside Norway (green box). \rr{Cross-border international transmission cables are not directly used \rr{in} the balancing market, but they affect the balancing through the day-ahead market and through \rr{unplanned changes in available transmission capacity.}} 

The consumption is highly correlated between zones as it is dominated by consumption with temporal patterns that are independent of market zones. The production is strongly correlated with consumption, although this is less clear for the consumption dominated zones (black box in \figref{fig:correlation_allzone_volumes}). On the contrary, we do not observe any significant correlation between balancing and production or consumption (orange box in \figref{fig:correlation_allzone_volumes}) \rr{confirming the role of the balancing as adjustments to unforeseen deviations}.

For the individual (dominating) production categories, we observe strong correlation between zones for the individual types of production: hydro reservoir, run-of-river and onshore wind. This shows the complementary application of the sources where for example hydro reservoir is used to compensate for over/under production of wind power. There are no significant correlations between the individual types of production and the balancing volume \rr{confirming the observations from \secref{sec:temporal_vol} of smaller increase in balancing volumes than increase in wind power production.}

\subsection{Market relations for prices}
It is known that coupling of markets leads to a price convergence between the coupled markets \citep{keppler_impacts_2016}. Due to transfer capacities between the zones, we expect a correlation between the volumes and prices of the Norwegian zones, in particular NO1--2--5 and NO3--4. This is clearly visible in the correlation heatmap in \figref{fig:correlation_allzones_prices} (in \appref{app:Appendix}) where we observe correlation between zones, also those that are not directly neighbouring. 

Regulation prices are strongly correlated with day-ahead prices and secondary with balancing volumes. 
The deviations from the day-ahead prices are somewhat correlated with the balancing volume, allowing for price predictions given you can predict the balancing volume.

Previous research considered the relation between price premiums (difference between day-ahead price and balancing price) and balancing. In particular \citet{klaeboe_benchmarking_2015} found a decreasing Pearson correlation between 2003--2007 (0.78) and 2010--2012 (0.47) by comparing with \citet{jaehnert_modelling_2009} with the caveat that NO1 was split into NO1 and NO2 between the two time intervals.
However, Pearson correlation only accounts for linear relationships. Instead we apply the Spearman correlation which evaluates the monotonicity of a relationship, leading to significantly higher values for all zones. \figref{fig:delta_correlations} shows the Pearson and Spearman correlation coefficients between balancing volumes and price premiums per year for NO1--5. We compare the values from the literature with those from 2016--2022. As the two methods incorporate the same range (-1 to 1) we loosely interpret the dropping Pearson correlation for NO1 as a decrease of linear relation and an increase in non-linear relations over time.

\section{Simple modelling and feature importance} \label{sec:model}
\rr{To further investigate relations in the data that may not be significant in the statistical analysis, we perform a simple modelling and feature importance analysis.}
The observed increased complexity motivates the use of advanced data driven methods over the classical statistical methods used in previous research for predicting balancing volumes a few hours ahead. Given the construction of the balancing market, it is per default difficult to predict multiple hours ahead. Here the focus is to obtain a model based on market data alone to investigate the importance of the various features. The challenges of a full scale predictive model are discussed in \secref{sec:features}.


\subsection{Model architechture, training and performance}
We train a \rr{snapshot model to predict balancing volumes ahead in time using training data for} balancing price and volume, day-ahead price and volume, hour of day, day of week, month and year for the Norwegian market zones. The model is configured to predict the balancing volumes for hour 0 based on information from previous hours. Consequently, in order to predict four hours ahead, we provide the balancing volumes for hour t-4, t-5 and t-6 as input. The day-ahead prices are given for hour $t_i$ for i $\in \{-3,...,3\}$ since these are settled at least 12 hours prior to realisation, and will always be available for prediction \rr{up to 12 hours ahead}. 

\rr{The balancing market is dominated by hours with no activity, and at the same time, the main interest of a predictive model is to capture the sparse events where the balancing volumes are high. This is a challenging scenario, which we work around by a naively resampling of the balancing data. We randomly sub-sampled} the hours with zero balancing such that they provided 1/3 of the total training sample, while 2/3 of the sample was hours with non-zero balancing volumes.

Hourly data for the period 2016--2022 was randomly shuffled into a training and a test set using a 0.7/0.3 split. The hyperparameters of the XGBoost algorithm were chosen as n\_estimators=400, max\_depth=15, eta=0.1, subsample=0.7 and colsample\_bytree=0.8, and mean squared error loss. Hyper-parameter tuning by trial and error showed that increasing n\_estimators and max\_depth from their default values increased the performance of the model significantly. 
The $R^2$ reported by XGBoost on the test data range from 0.47 to 0.57 across the different bidding zones as given in \tabref{table:feature-importances}.

\subsection{Feature importance} \label{sec:features}
We assess the predictive power of different features by \rr{SAGE values}, which estimates how much each feature contributes to the model's predictive power\rr{\footnote{SAGE values are not designed for correlated data, and hence the values may include some uncertainty and small differences should be interpreted with caution.}}. The results are given in \tabref{table:feature-importances}. The overall model performance is poor, highlighting the challenging nature of the problem. We observe that the features with the highest predictive power are the \rr{temporal} lags of the balancing volume from the corresponding bidding zone. This is expected given the high auto-correlation of the first lags of balancing volumes. 

Since the balancing market is designed to account for deviations between planned production and the actual demand, the model has potential for improvement by adding features that carries information about the origin of those deviations such as changes in weather or unforeseen changes in the infrastructure. Simple attempts on including representative temperatures or wind speed/wind direction did not improve the model, and a full study is beyond the scope of this paper where we focus on the market data. Based on the large auto-correlation with one hour lag, a model framework such as recurrent neural networks should be considered. In addition, changing societal factors such as power production policies may affect such modelling, and the framework should allow for continuous updates with new data. \rr{We conclude that public available power market data does not provide necessary information for predicting balancing volumes.}

\begin{table*}[hbtp]
\centering
\begin{tabular}{c c|c|c|c|c} 
  & \textbf{NO1} ($R^2$=0.50) & \textbf{NO2} ($R^2$=0.49) & \textbf{NO3} ($R^2$=0.56) & \textbf{NO4} ($R^2$=0.57) & \textbf{NO5} ($R^2$=0.47) \\ 
 \hline
 \hline
 1 & $NO1_{imb-volume}^{-4}$ (1)    & $NO2_{imb-volume}^{-4}$ (1)    & $NO3_{imb-volume}^{-4}$ (1)    & $NO4_{imb-volume}^{-4}$ (1)     & $NO5_{imb-volume}^{-4}$ (1)\\ 
 2 & $NO5_{imb-volume}^{-4}$ (0.15) & $NO2_{imb-volume}^{-6}$ (0.14) & $NO4_{imb-volume}^{-4}$ (0.08) & year (0.4)         & $NO2_{imb-volume}^{-4}$ (0.25) \\ 
 3 & $NO2_{imb-volume}^{-4}$ (0.14) & $NO5_{imb-volume}^{-4}$ (0.07) & $NO3_{imb-volume}^{-6}$ (0.06) & $NO4_{imb-volume}^{-5}$ (0.15)  & $NO5_{imb-volume}^{-6}$ (0.13) \\ 
 4 & $NO1_{prod-volume}^{-0}$ (0.09)    & $NO2_{imb-volume}^{-5}$ (0.06)       & $NO3_{imb-volume}^{-6}$ (0.05)       & $NO4_{imb-volume}^{-6}$ (0.15)  & month (0.13) \\ 
 5 & $NO1_{imb-volume}^{-6}$ (0.08) & $NO1_{imb-volume}^{-4}$ (0.05) & $NO3_{prod-volume}^{-0}$ (0.05)    & $NO1_{DA-price}^{-0}$ (0.11)     & $NO5_{imb-volume}^{-5}$ (0.11)\\ 
\end{tabular}
\caption{List of the five most important features in the XGBoost regression model for determining the balancing volumes in NO1--5. The subscript depicts the feature type and the superscript depicts the time-lag of the data. The coefficient of determination ($R^2$) evaluated on the test data is given for each model. The value in parenthesis is the relative importance based on SAGE values expressed as a fraction between 0 and 1. Sell and buy refers to the volumes traded in day-ahead, imb-volume is the imbalance volume while DA-price refers to the day-ahead price.}
\label{table:feature-importances}
\end{table*}

\section{Results and discussion} \label{sec:results}

Our analysis shows that the balancing market is strongly coupled to the day-ahead market. In addition, infrastructure, policy and behavioural changes affecting the day-ahead market will also affect the balancing market. This applies for example to a growing degree of electrification, increase of volatile production from intermittent energy sources such as wind and solar, and the geopolitical development having an strong impact on overall energy prices.

When investigating temporal evolution, it becomes clear that there is no such thing as a standard year in the power market. Given the correlations to other zones and countries, the Norwegian zones are affected by international events and policy changes e.g. related to supply stability. This will affect the volumes directly, while the balancing prices are affected both by the balancing volumes and by policies for price setting. It is also observed that increased volatile production does not increase the number of hours with regulation, and we interpret that situations that were difficult to predict before, remains difficult to predict \rr{presently and quite potentially in the future}. However, increasing volumes makes the impact of wrong predictions larger.

\rr{International connections are not used directly in the regulation market, but they affect the day-ahead market and the bilateral trading in the intra-day markets which again can affect the regulation indirectly. It is possible that the introduction of interconnectors affects the strategies chosen by the traders, and that this in turn gives new balancing needs. There is also a question of strategy for each trader or trading organisation as to which market to settle imbalances in. If sufficient lead time is given, a producer may trade in the intra-day market to secure the fulfilment of commitments made in the day-ahead market that have become hard to fulfil based on intended production assets. We do not have access to insight in the individual trading strategies, but it is expected that each trading organisation has a trading strategy for these situations. It is reasonable to assume that the strategy is developing and adapting to the observed changes in the power system. Hence, the risk-handling strategy and other trader specific choices are expected to impact the resulting prices and volumes in the balancing markets. Such input is beyond the scope of this paper and will only be captured by the statistical analysis as irregularities and non-stationarity.}

\rr{The introduction of more non-regulated power production such as wind and solar power is in and of itself expected to give larger volumes and more hours of non-zero regulation in the regulatory markets. It is observed in this study (for example in \figref{fig:ratio_evolution}) that the resulting behaviour in the regulation market is more nuanced. It is not found that the number of hours with non-zero regulation volumes have increased significantly, rather the opposite has been found as a trend for several zones. Although the volumes have increased for those non-zero hours. In addition, the increase in the ratio of non-regulated power production does not seem to correlate with the increase in the regulation volumes. A potential mechanism for these observations may be that the balancing strategies of the producers have developed, for example by building larger portfolios of diverse assets that combined give better predictability. As noted, the statistical analysis shows how the market is developing. The fact that this residual market shows stochasticity indicates that the day-ahead market is well-functioning and interacts with the balancing market in an efficient manner. With public data, the analysis could be performed by any entity ensuring market transparency. This is also valid for other markets where similar data is available.}

Due to the demonstrated changing characteristics of the balancing market, modelling of this is very challenging and will require methodology that is able to handle systems in development \rr{displaying non-linear behaviour}. It is a strong recommendation that such modelling is performed in a way that it may be continuously updated with new data from the system in order to capture the ever changing characteristic of the system.


\section{Acknowledgment}
This work was supported by Research Council of Norway (RCN) grant number 309315. The KoBas project is co-founded by the Norwegian Research Council, with Skagerak Kraft, Eviny Fornybar, Equinor and SINTEF as partners.

\appendix
\section{} \label{app:Appendix}
\begin{table}[!h]
    \begin{tabular}{l|r|r|r|r|r}
                        & NO1     & NO2      & NO3      & NO4     & NO5      \\ \hline \hline
    {\bf Regulation}    & & & & & \\ 
    {\bf volume}    & & & & & \\ 

    Median [MWh]                 & 0.0    & 0.0     & 0.0     & 0.0    & 0.0   \\ 
    Mean [MWh]                 & -5.0    & -16.1    & -34.9    & -21.5   & -19.5   \\ 
    Std [MWh]              & 35.2    & 140.6    & 112.9    & 112.7   & 175.7   \\ 
    $\%$ zero [-]                & 69      & 47       & 44       & 59      & 41    \\ \hline
    {\bf Up price }      & & & & & \\ 
    {\bf difference}      & & & & & \\ 

    Median [€/MWh]                & 0.0    & 0.0     & 0.0     & 0.0    & 0.0   \\ 
    Mean [€/MWh]              & 2.6    & 2.3     & 2.81    & 2.29   & 2.55   \\ 
    Std [€/MWh]           & 12.7   & 10.6    & 42.3    & 41.2   & 11.7   \\ 
    $\%$ zero                & 69     & 69      & 71      & 74     & 69    \\ \hline
    {\bf Down price }    & & & & & \\ 
    {\bf difference}    & & & & & \\ 
    Median [€/MWh]                & 0.0    & 0.0     & 0.0     & 0.0    & 0.0   \\ 
    Mean [€/MWh]              & -5.4   & -5.6    & -3.6    & -2.9   & -5.3   \\ 
    Std [€/MWh]           & 20.8   & 21.4    & 9.5     & 8.2   & 20.6   \\ 
    $\%$ zero                & 55     & 55      & 52      & 56     & 56    \\
    \end{tabular}
    \caption{Median, mean, standard deviation and percentage of hours with zero, for balancing volume and price differences for NO1--NO5 for 2016--2022. Volumes are in MWh and prices in €/MWh.}
    \label{tab:global_stats}
\end{table}

\begin{figure*}
    \centering
    \includegraphics[width=0.9\linewidth]{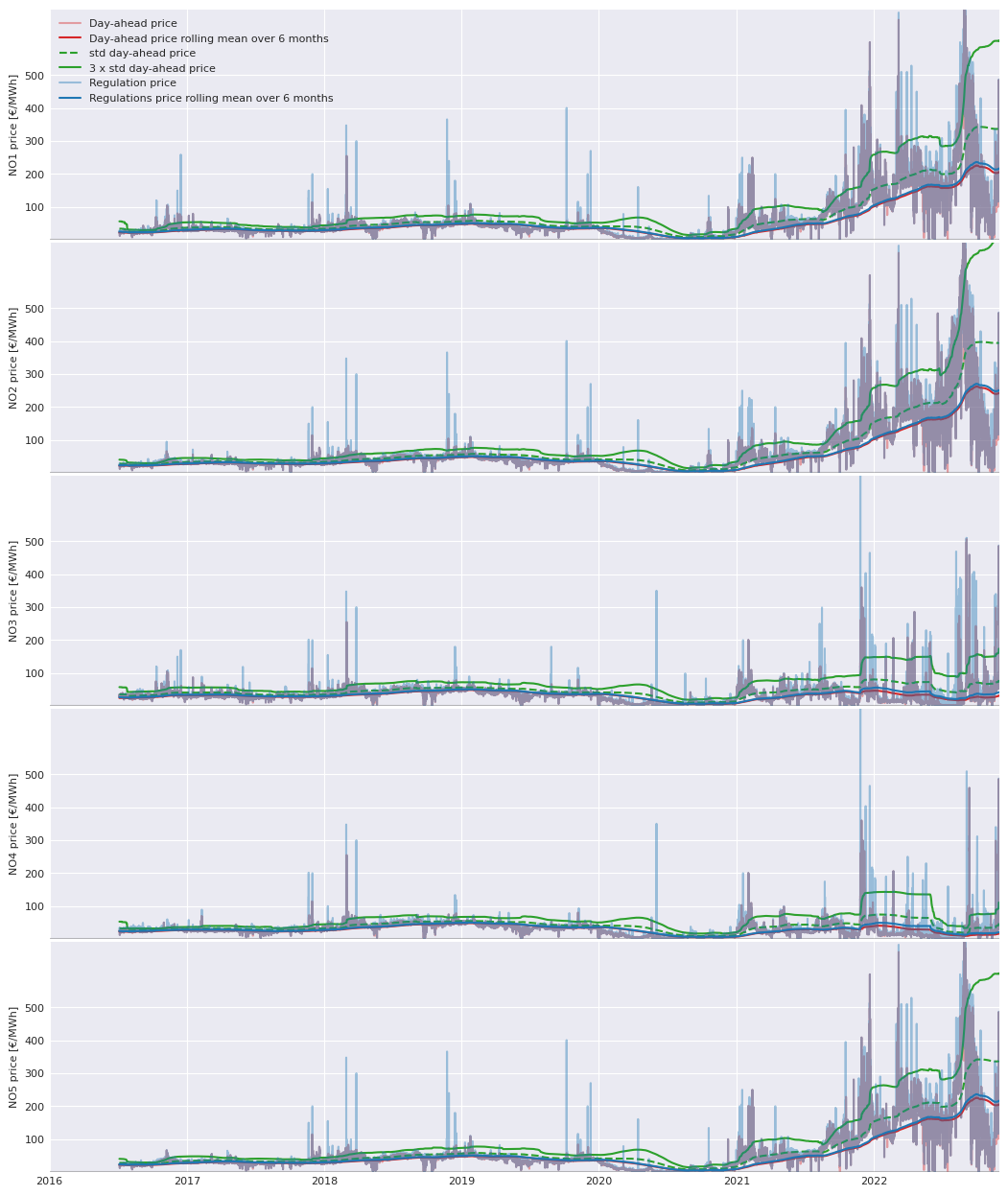}
    \caption{Per hour day-ahead and balancing prices for together with the six-month moving average and standard deviation for balancing price calculated using one hour incremental rolling (backwards in time) from 2016 to November 2022 for NO1 (top) to NO5 (bottom). Due to large transmission capacity between zones, NO2 and NO5 are similar to NO1, while NO4 is similar to NO3. A few counts with prices higher than 700 €/MWh are outside the plotting range}
    \label{fig:regulation_price_moving_avg_full}
\end{figure*}

\begin{figure*} [b]
    \centering
    \includegraphics[width=\textwidth]{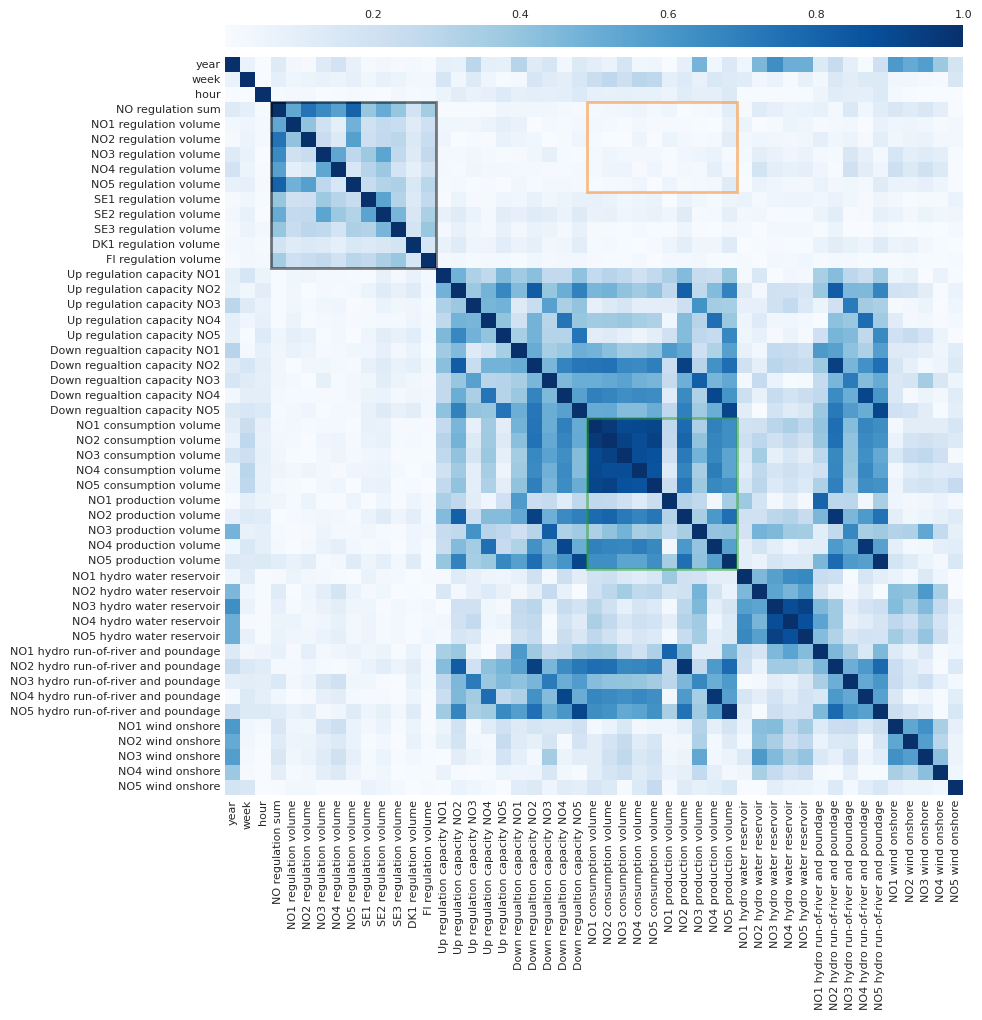}
    \caption{Spearman correlation heat map for the period 2016--2022. Correlations between balancing volumes (regulation) of the Nordic market zones plus neighbouring zones (black box), consumption and production volumes (green and orange boxes), volumes from various production types and available balancing capacity.}
    \label{fig:correlation_allzone_volumes}
\end{figure*}

\begin{figure*}
    \centering
    \includegraphics[width=\textwidth]{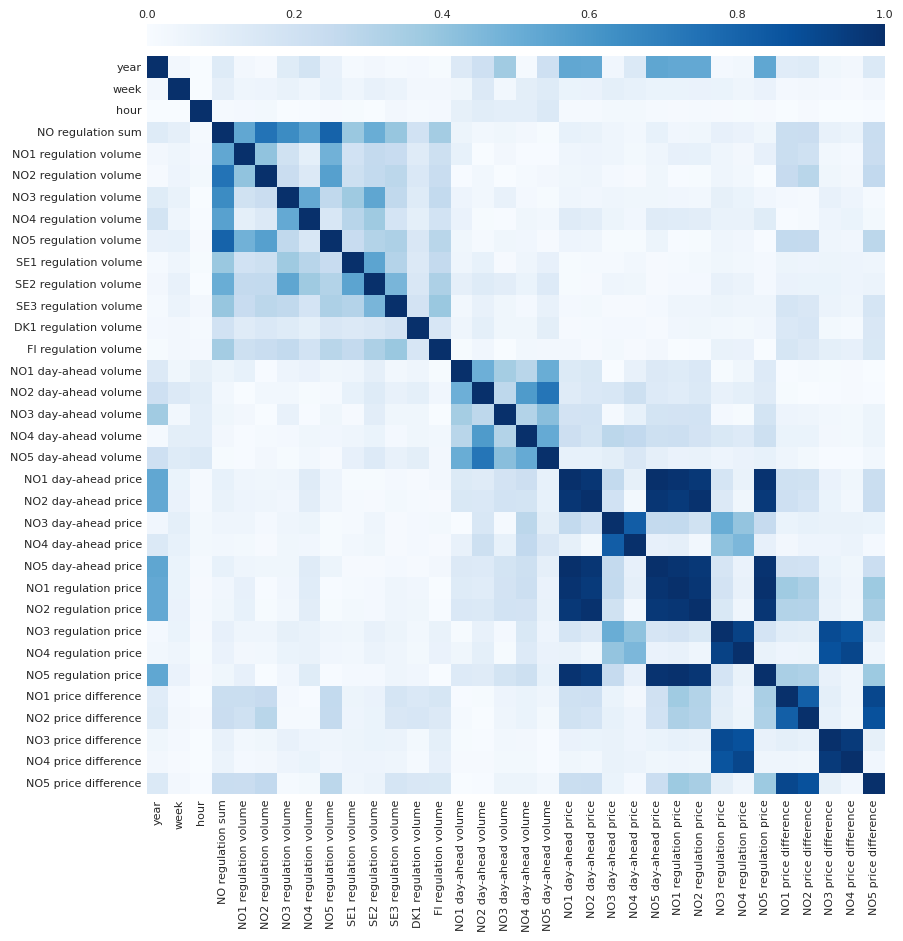}
    \caption{Spearman correlation heat map for the period 2016--2022 for balancing volumes (regulation), day-ahead prices and price differences.}
    \label{fig:correlation_allzones_prices}
\end{figure*}

\begin{figure*}
    \centering
    \includegraphics[width=0.9\linewidth]{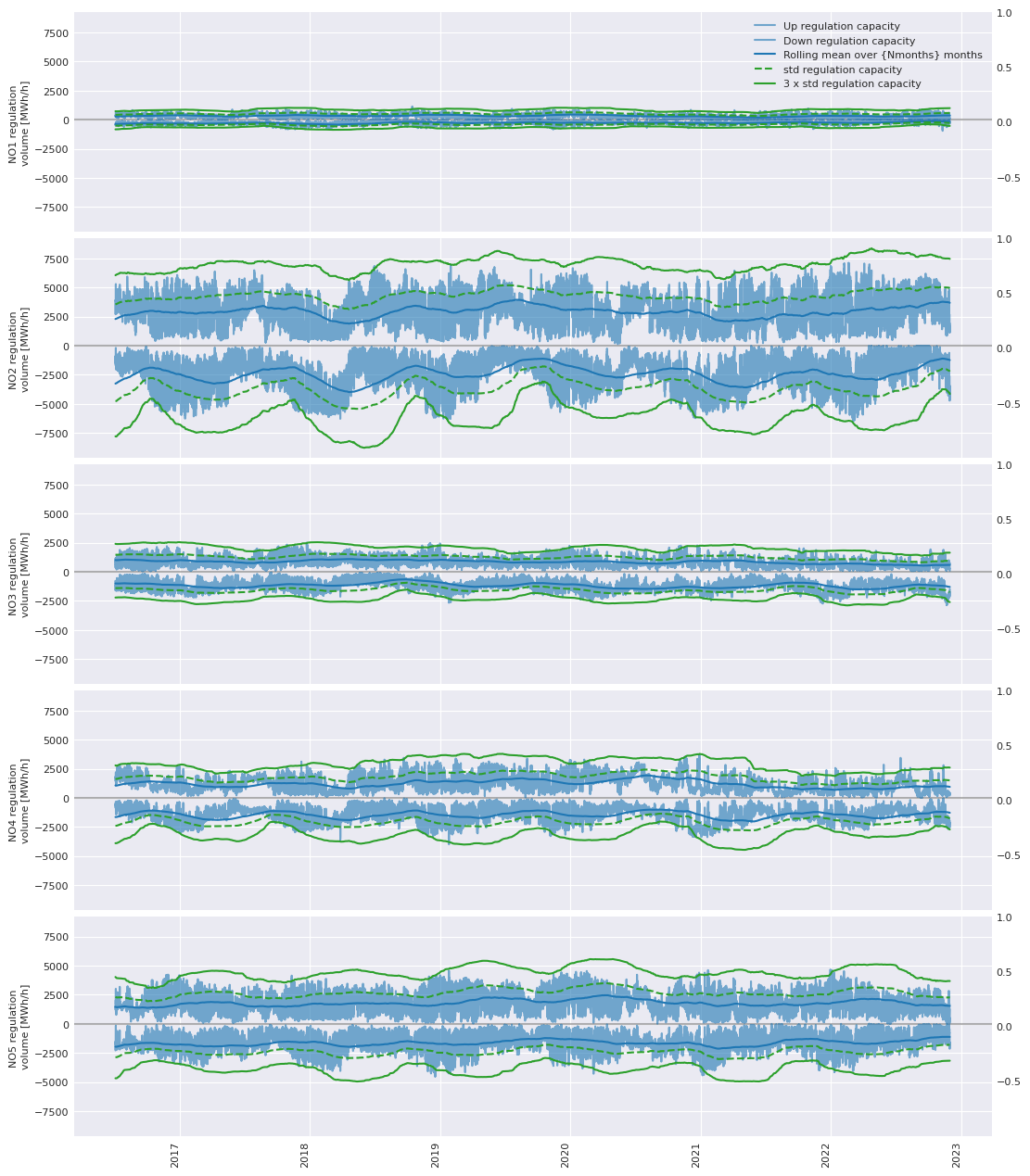}
    \caption{Available balancing capacity volumes for up- and down-regulation (faint blue, positive and negative values respectively), and six-months moving average (blue line) and standard deviation (green) for one hour incremental rolling for 2016--2022 for NO1 (top) to NO5 (bottom). Positive values correspond to available up-regulation and negative to available down-regulation. All zones show some seasonal variation in regulation capacity. NO2 and NO5 dominates the regulation capacity due to large amounts of flexible hydro reservoir power in the zones.}
    \label{fig:regulation_capacity_moving_avg}
\end{figure*}

\twocolumn